\documentclass[doublecol]{epl2}

\usepackage{amsmath,amsfonts,amssymb,bm,slashed,bbm}%,mathrsfs}
\usepackage{graphicx,color}
\renewcommand{\Ref}[1]{(\ref{#1})}
\newcommand{\eq}[2]{\begin{align}\label{#1}#2\end{align}}

\newcommand{\nn}{\nonumber}
\renewcommand{\ni}{\noindent}
\newcommand{\pa}{\partial}

\newcommand{\al}{\alpha}
\newcommand{\ga}{\gamma}
\newcommand{\la}{\lambda}
\newcommand{\om}{\omega}\newcommand{\Om}{\Omega}
\newcommand{\vphi}{\varphi}

\title{Casimir effect for scalar field rotating on a disk}
\shorttitle{Casimir effect for scalar field rotating on a disk}

\author{M. Bordag\inst{1,2}\footnote{bordag@theor.jinr.ru} \and  I.G. Pirozhenko\inst{1,3}\footnote{pirozhen@theor.jinr.ru}}
	
\institute{
\inst{1}	Bogoljubov Laboratory of Theoretical Physics, Joint Institute for Nuclear Research,141980 Dubna, Russian Federation\\
\inst{2}Institute for Theoretical Physics, University Leipzig,	IPF 231101, D-04081 Leipzig, Germany\\
\inst{3} 	Dubna State University,  Universitetskaya str., 19,  141986 Dubna, Russian Federation
}

\abstract{
We compute the vacuum energy of a scalar field rotating with angular velocity $\Om$ on a disk of radius $R$ and with Dirichlet boundary conditions. The rotation is introduced by a metric obtained by a Galilean transformation from a rest frame. The constraint $\Om R<c$ must be obeyed to maintain causality. To compute the vacuum energy, we use an imaginary frequency representation and the well-known uniform asymptotic expansion of the Bessel function.  We use the zeta-functional regularization and separate the divergent contributions, which we discuss in terms of the heat kernel coefficients. The divergences are found to be independent of rotation. The renormalized finite part of the vacuum energy is negative   and becomes more negative for larger rotation frequencies. 	 	
}

\begin{document}
\maketitle
%\tableofcontents

\section{\label{T1}Introduction}	
Vacuum energy, especially its manifestation in the Casimir effect, is a fundamental part of our understanding of the quantum world and its investigation reveals many interesting features. A relatively new area of application are rotating systems. With respect to the vacuum energy these are not yet well studied. It should be mentioned, that the topic of rotating fields has gained interest from the ongoing and planned experiments of heavy ion collisions where, a rotating plasma, and with it a system of rotating hadronic fields, is formed  for a short time. A more conventional topic is rotating fields in gravity, e.g., Kerr solutions. But these are classical.

So we take our motivation from heavy ion collisions and study relativistic quantum fields in a rotating frame of  reference. In this approach, the rotation is given by the metric tensor, or equivalently, by the interval. With such background, one must consider the Klein-Gordon equation, to define the vacuum and can go on to calculate quantum effects such as the vacuum energy. 

The vast majority of investigations of rotating fields work with a frame which is set in rotation by a coordinate transformation,
% Galilean transformation \textcolor{black}{of the angular variable $\vphi$},
%
\eq{1.1}{t'=t,\ \vphi'=\vphi-\Om t, \  r'=r, 
}
with the angular velocity $\Om$. Such frame is rigidly rotating \textcolor{black}{and space is limited to}
\eq{1.2}{ r <\frac{c}{|\Om|}
}
by causality. While this restriction is not a problem in non-relativistic physics, e.g., for a rotating fluid on Earth, for a relativistic field $\phi(t,\vec x)$, which is in principle defined in all space, one is forced to introduce \textcolor{black}{artificially} a limited volume. Usually, a cylinder around the axis of rotation, with a radius $R$ obeying $R<c/\Om$, is taken. In order to keep the operators self-adjoint, a corresponding boundary condition must be imposed. The price for  keeping causality is to answer the question to what extent this artificial boundary affects the properties of the rotating field one is interested in. An advantage of the transformation \Ref{1.1} is that despite of the acceleration, the vacuum is the same in the resting and in the rotating frame.

For the  restriction \Ref{1.2}, there is a known way out by considering a relativistic generalization of \Ref{1.1},
%Lorentz transformation \textcolor{black}{for the angular variable}, %*** deepwrite bis hier
%
\eq{1.3}{t'&= t \cosh(\theta)-\frac{r\vphi}{c}\sinh(\theta), \ 
	\vphi'= \vphi \cosh(\theta)-\frac{ct}{r}\sinh(\theta), \nonumber\\
&	\ r'=r, 
}
with $\theta=\frac{\Om r}{c}$, to set the frame into rotation. This transformation was first considered as early as \cite{fran22-8-265} (and rediscovered several times\textcolor{black}{, for a more recent discussion see \cite{delo00-17-4241}}).  There are no problems with causality \textcolor{black}{and no restriction like \Ref{1.2} is necessary}. However, this transformation raised a large number of discussions,  especially concerning the definition of the vacu\-um and the synchronization of clocks when going round ($\vphi\to\vphi+2\pi$), which we will not touch here.  It has never been considered in the context of vacuum energy.

Vacuum energy is known to be influenced by a gravitational background like any other inertial mass, \cite{full09-42-155402}. Of course, this is also true for the metric  $g_{\mu\nu}$, \Ref{2.6}, which appears in a rotating frame with \Ref{1.1}. Thus the  effects of rotation were studied in \cite{cher1203.6588} and \cite{cher13-87-025021} (proposing a 'perpetuum mobile of the fourth kind') for a Dirichtet point rotating on an $S^1$. A similar investigation was done on the lattice for QCD, revealing a negative moment of inertia, \cite{brag24-110-014511} and \cite{brag24-852-138604}.

With the rotation given by the Galilean transformation \Ref{1.1}, it is known that the one particle energy can be lowered by a, say, scalar field. In the last decade there have been a number of investigations of a rotating scalar field where this decay is enhanced by a magnetic field parallel to the axis of rotation and an instability of the ground state with formation of a condensate may appear, \cite{liuz18-120-032001} and \cite{guot22-106-094010}. 
We also mention the investigation of the Casimir effect in the background of a Kerr black hole in \cite{nour10-82-044047} and of a rotating sphere between the plates, \cite{beze14-89-044015}.

%%%%%%%%%%%%%%%%%%%%%%%%%%%%%%%%%%%%%%%%%%%%%%%%%%%%%%%%%%%%%%%%%%%%%%%%%%%%%%%%%%%%%%%%%%%%%%%%

In the present paper we consider the Casimir effect for a massive scalar field rotating on a disk with Dirichlet boundary conditions. We restrict ourselves to a disk since this is the simplest case where the  features  specific to rotation can be studies. We use the formalism developed a quarter of a century ago and applied to a variety of settings, \cite{BKMM} and \cite{bord01-353-1}, and develop the necessary modifications. An interesting question to be answered is the dependence of the sign of the vacuum energy on the geometry and other given conditions, including the question of whether a negative vacuum energy can overcompensate for the classical rotational energy. Another question is the validity of the approach based on the Galilean transformation \Ref{1.1} for a relativistic field, e.g. will a slow rotation be a valid approximation?\smallskip

\ni Throughout the paper we use units such that $c=\hbar=1$.

\section{General formulas}
We consider a real scalar field on a disk of radius $R$ in the metric 
\eq{2.1}{  g_{\mu\nu}=
	\left(\begin{array}{ccc}
		1- v_1^2-v_2^2& -v_1 & -v_2\\
		-v_1 & -1 &0\\
		-v_2 & 0 & -1	
		 \end{array}\right),
}
where $\vec v=\Om r\left(\begin{array}{c}-\sin(\varphi)\\\cos(\varphi)\\0 \end{array}\right)$ is the speed of a rotating point. The corresponding Klein-Gordon equation reads
\eq{2.2}{  \left(  (\pa_0-\vec v \vec \nabla)^2-\Delta+m^2\right)\phi(x)=0.
}
The metric \Ref{1.1} results from a Galilean transform of coordinates and only points inside a disk of radius $R$ with $\Om R<1$ do not exceed the speed of light. For this reason, it is common to restrict the field to such disk and to demand boundary conditions. We take Dirichlet ones.

Separating the variables, the solution of equation \Ref{2.2} can be written in the form,
\eq{2.3}{ \phi(x)=\sum_{l=-\infty}^\infty e^{-i\om_{l,n}t+i l \varphi}J_l(\textcolor{black}{\al}  r),
}
where $J_l(z)$ is a Bessel function of first kind. The Dirichlet boundary condition defines $\textcolor{black}{\al}$ in terms of the zeros $j_{l,n}$ of this Bessel function, $\textcolor{black}{\al}=\frac{j_{l,n}}{R}$, and the eigenfrequencies can be written down immediately,
\eq{2.4}{ \om_{l,n} =-\Om l+\sqrt{m^2+\left(\frac{j_{l,n}}{R}\right)^2},
}
where $j_{l,n}$ are the zeros of the Bessel function $J_l(z)$. 
From the spectrum \Ref{2.4}, we get the vacuum energy $E_0$ using zetafunctional regularization  in the form
\eq{2.5}{ E_0 = \frac{\mu^{2s}}{2} \sum_{l=-\infty}^\infty \sum_{n=1}^\infty
	\om_{l,n}^{1-2s},
}
where $s$ is the  regularization parameter, $s\to 0$ at the end, and $\mu$ is the dimensional parameter associated  with this regularization.

\begin{figure}[t]
	%file is C:\\Users\\bordag\\WORK\\Physics\\\Vacuum, effPot, \
	%Casimir25\\Casimir25\\250314 partB.nb
	\includegraphics[width=0.40\textwidth]{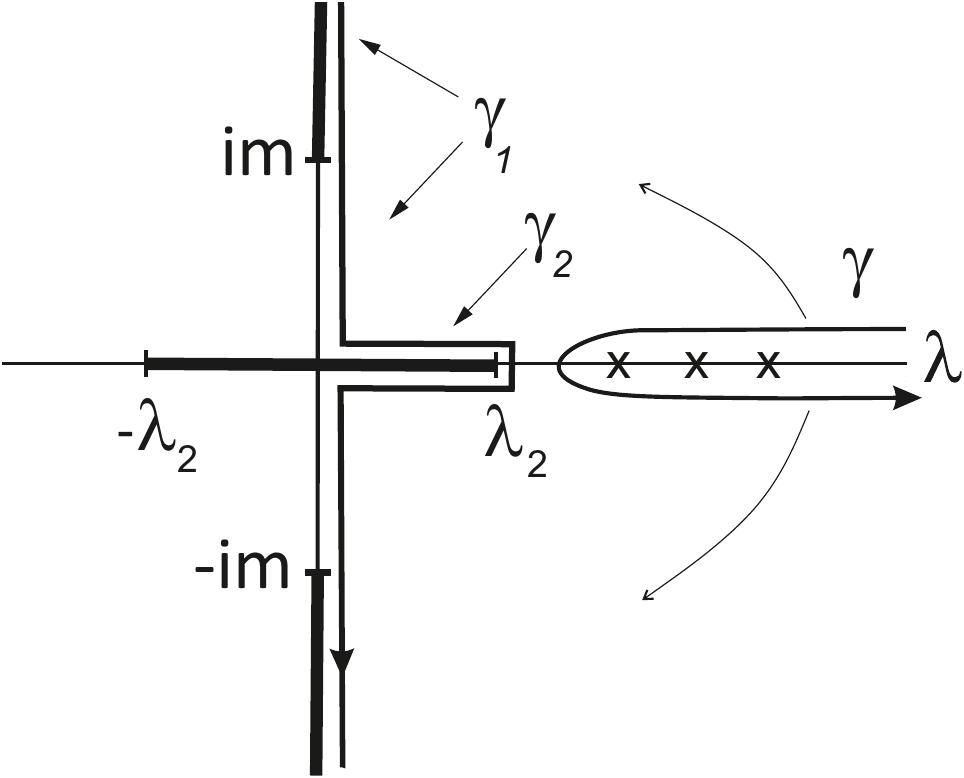}
	\caption{
		The integration contour for \eqref{2.7}, $R=1$, $\Omega l>m$. The crosses mark the zeros $j_{l,n}$ of the mode generating function \Ref{2.6}. See the text for more details.}		\label{fig:fig1}	
\end{figure}

The problem to be solved is the analytic continuation of $E_0$ to $s=0$. One way is to use the asymptotic behavior of the frequencies $\om_{l,n}$ for large indices. This is only useful for simple problems. Instead, we proceed by transforming the sum over $n$ into an integral using 
\eq{2.6}{ \Phi(\la)=\la^{-l} J_l({\la} )
}
as the mode generating function, $\Phi(j_{l,n})=0$, which is regular at $\la=0$.
We get
\eq{2.7}{ E_0&=\frac12\sum_{l=-\infty}^\infty
	\int_\gamma\frac{d\la}{2\pi i} \, g\left({\la}/{R}\right)^{1-2s} \frac{\pa}{\pa\la}\ln\Phi(\la), \\
	&g(\lambda)= -\Om l+\sqrt{m^2+\la^2},
	\label{2.7a}
}
where the integration path $\ga$ surrounds the real positive zeros of  the mode generating function  $\Phi(\la)$, see fig.~\ref{fig:fig1}. 

The contour $\gamma$ encloses the positive real zeros of the mode generating function \eqref{2.6}. 
Further, in the complex plane there are two branching points, $\pm \lambda_2$, given by 
\eq{2.7b}{\lambda_2 =\sqrt{(\Omega l)^2-m^2}.
}
These are on the real axis when $\Omega l>m$, and exist only for $l>0$. There is another pair of branch points on the imaginary axis, $\pm i m R$, which disappear for $m=0$.

With $E_0$, eq. \Ref{2.7}, we have a representation for the regularized vacuum energy. It is, in the given regularization, a meromorphic function of $s$ with a pole at $s=0$, which results from the ultraviolet divergence. It is convenient to study the divergences in terms of the heat kernel coefficients. For $2+1$ dimensions, the heat kernel and its expansion for $t\to 0$ are
\eq{h.1}{K(t)=\sum_{l=-\infty}^\infty\sum_{n=1}^\infty
	e^{-t(\om_{l,n}^2+m^2)}
	\simeq\frac{1}{4\pi t}\sum_{k\ge 0}a_kt^k e^{-tm^2}.
}
In terms of the heat kernel coefficients, $a_k$, the divergent part of the vacuum energy is 
\eq{h.2}{E_0^{div}=\frac{m^2a_{1/2}-a_{3/2}}{16\pi^{3/2}s}
	+O(1)
}
and can be derived in parallel to equation (4.30) in \cite{BKMM}.  Equation \Ref{h.2} is the semiclassical expansion of the vacuum energy for large mass and contains the non-negative powers of $m$. It must be subtracted from the regularized energy $E_0$, \Ref{2.7} to obtain the renormalized one,
\eq{h.3}{E_0^{ren}=E_0-E_0^{div}
}%see 250312 heat kernel.nb
at $s=0$. In the massless case there is only the coefficient $a_{3/2}$ (which replaces $a_2$ in (3+1) dimensions) to subtract, and the result contains the logarithm of the dimensional factor $\mu$ (which we did not note in intermediate steps). We will discuss this in the Conclusions.

For simplicity, we will restrict ourselves to the massless case, m=0.
Then the branch points $\la_2$, \Ref{2.7b}, and the corresponding section are only on the real axis. Further, we put $R=1$. The dimension can be restored by substituting $E_0\to E_0/R$ and $\Om\to \Om R$. Also, we assume $\Om>0$.

\section{Calculation of the vacuum energy}
In this section we calculate the regularized vacuum energy using the formulas \Ref{2.7} which was prepared in the preceding section.
For these calculations, it is meaningful to deform the integration contour moving it towards the imaginary axis, as shown in figure \ref{fig:fig1}.  
It is convenient to consider the contributions from $l=0$ and from the paths $\ga_1$ and $\ga_2$ separately and we split the vacuum energy correspondingly into three parts,
\eq{2.8}{ E_0=E_A+E_B+E_{l=0},
}
where $E_A$ results from the vertical part of the path and  $E_B$ from the horizontal one.

\subsection{Part A of the vacuum energy, $E_A$}
In this section we calculate part A of the vacuum energy, defined in \Ref{2.8}. It takes the form
\eq{3.1}{ E_A&=\sum_{
			l=-\infty\atop l\ne0
	}^\infty \int_0^\infty \frac{d\xi}{4\pi i}
	\left[g(-i \xi)^{1-2s} - g(i \xi)^{1-2s}     \right]
		 \frac{\pa}{\pa\xi}\ln\Phi(i\xi),
}
where $g(i\xi)$ is defined by \eqref{2.7a} with $m=0$, and  we accounted for the reality of the function $\Phi(i\xi)$. Performing the analytic continuation, one arrives at
\eq{3.3}{ E_A =-\frac{1}{2\pi}\sum_{l=1}^\infty \int_0^\infty d\xi
	\left((\Om l)^2+\xi^2\right)^{\frac12-s} h(\xi/l) \ 
	 \frac{\pa}{\pa\xi}\ln\Phi(i\xi)
}
with
\eq{3.4}{ h\left(\frac{\xi}{l}\right)&=\sin\left[(1-2s)
	\left(\frac{\pi}{2}
	-\arctan\left(\frac{\Om l}{\xi}\right)\right)\right]
 \\\nn&~~	+\sin\left[(1-2s)\left(\frac{\pi}{2}
	+\arctan\left(\frac{\Om l}{\xi}\right)\right)\right].
}
We proceed with the substitution $\xi=l z$ and note
\eq{3.5}{ E_A =-\frac{1}{2\pi}\sum_{l=1}^\infty l^{1-2s}
	\int_0^\infty dz
	\left(\Om^2+z^2\right)^{\frac12-s} h(z) \ w(z),
}
where we defined
\eq{3.6}{ w(z)=	\frac{\pa}{\pa z}\ln\left(z^{-l}I_l(l z)\right)
}
with the modified Bessel function  $I_l(l z)$.

To do the continuation in $s$, we subtract and add the function $w^{as}(z)$,
\eq{3.7}{w^{as}(z)=\sum_{k=-1}^2\frac{w^{as}_k(z)}{l^k},
}	
which is defined as part of the expansion of $w(z)$ for large $l$ up to $l^{-2}$. It can be obtained from the uniform asymptotic expansion of the modified Bessel function,
\eq{3.8}{ I_l(l z)= e^{l\eta(p)}\sqrt{\frac{p}{2\pi l}}
	\left(\sum_{k=0}^{\infty}\frac{U_k(p)}{l^k}\right),  
}
$$ p=\frac{1}{\sqrt{1+z^2}}, \quad \eta(q)=\frac{1}{p}+\ln\frac{z}{1+p^{-1}},$$
where the coefficients $U_k(p)$, can be found, for example, in \cite{AbramowitzStegun2010} and \cite{NIST:DLMF}, eq. (10.41.10) 

With \Ref{3.7} we represent
\eq{3.9}{ E_A=E_A^{num}+E_A^{as}
}
with
\eq{3.10}{ E_A^{num} &=-\frac{1}{2\pi}\sum_{l=1}^\infty l^{1-2s} \int_0^\infty dz
	\left(\Om^2+z^2\right)^{\frac12-s}
	\\\nn&~~~~~~~~~~~~~~~~~~~~~~~~~~~~~~~~~~~\cdot h(z) \ (w(z)-w^{as}(z)),
	\\\nn E_A^{as}&= -\frac{1}{2\pi}\sum_{l=1}^\infty l^{1-2s} 
	\int_0^\infty dz \left(\Om^2+z^2\right)^{\frac12-s} h(z)w^{as}(z).
}
In the numerical part, $E^{num}_A$, the integration over $z$ converges for $s=0$. The resulting integrals decrease   sufficiently fast with growing  $l$, and the sum also converges for $s=0$. Using the property 
\eq{3.11a}{{h(z)_|}_{s=0}=\frac{2z}{\sqrt{\Om^2+z^2}}
}
and performing the integration numerically, we get
\eq{3.11}{E_A^{num}&=
	-\frac{1}{\pi}\sum_{l=1}^\infty l^{1-2s} \int_0^\infty dz
	 \,z \ (w(z)-w^{as}(z))
	\\\nn&\simeq -0.0021858.
}%see file 250316_2 Rechecking.nb
This part of the energy is independent of the rotation, which can be guessed from \Ref{2.7} by formally setting $s=0$ and considering the sign of $l$.

The 'asymptotic' part, $E_A^{as}$, defined in \Ref{3.10}, can be rewritten with \Ref{3.7} in the form
\eq{3.12}{ E_A^{as}=\sum_{k=-1}^2E^{as}_{A,k},
}
with
\eq{3.13}{E^{as}_{A,k} = -\frac{1}{2\pi}\sum_{l=1}^\infty l^{1-2s-k}
	\int_0^\infty dz \left(\Om^2+z^2\right)^{\frac12-s} h(z)w^{as}_k(z).
}
We consider the integration over $z$. Its convergence depends on the behavior of $w^{as}_k$ for $z\to\infty$. From the coefficients $U_k(p)$ in \Ref{3.8} the behavior
\eq{3.13a}{ w^{as}_k(z)\sim\frac{1}{z^{k+1}}
}
can be obtained.
Thus, for $k=2$, the integration over $z$ converges for $s=0$ and results in
\eq{3.13b}{ \int_0^\infty dz\, z\, w^{as}_2(z)=\frac{\pi}{128},
}
where \Ref{3.11a} was used. Together with the zeta function in front we arrive at
\eq{3.13c}{E^{as}_{A,2} = -\frac{1}{512s}-\frac{\ga}{256}+O(s),
}
where $\ga$ is Euler's gamma.

For the contribution from $k=-1,0,1$ in \Ref{3.16}, the integration over $z$ does not converge for $s=0$. To proceed, in its integrand we subtract and add an asymptotic part, now of the function $h(z)$ (this is in addition to \Ref{3.9}), with
\eq{3.14}{h^{as}(z)=h_0+\frac{h_1}{1+z^2},
}
where $h_0$ and $h_1$ are defined by the expansion
\eq{3.15}{ h(z)\simeq h_0+\frac{h_1}{z^2}+\dots \qquad \mbox{for}
	\quad z\to\infty.
}
Here
\eq{3.15a}{ h_0=2 \cos(\pi s), \quad h_1=-\Omega^2(1-2 s)^2 \cos(\pi s).
}
Note that the special values $h_0=2$ and $h_1=-\Om^2$ of these coefficients for $s=0$. So we split, for $k=-1,0,1$, 
\eq{3.16}{E^{as}_{A,k} =H^{num}_k+H^{as}_k,
}
where we defined,
\eq{3.17}{H^{num}_k &=-\frac{1}{2\pi}\sum_{k=1}^\infty l^{1-2s-k}
	%\zeta(k+2s-1)
	\int_0^\infty dz \left(\Om^2+z^2\right)^{\frac12-s}  \\\nn &~~~~~~~~~~~~~~~~~~~~~~~~~~~~~~~~~~\cdot(h(z)-h^{as}(z))w^{as}_k(z),
	\\\nn H^{as}_k &=-\frac{1}{2\pi}\zeta(k+2s-1)
	\int_0^\infty dz \left(\Om^2+z^2\right)^{\frac12-s} 
	\\\nn&~~~~~~~~~~~~~~~~~~~~~~~~~~~~~~~~~~~~~~~~~~~~\cdot h^{as}(z)w^{as}_k(z),
}
In $H^{num}_k$ we may put $s=0$ since there is no divergence. The integration can be done numerically and the summation results in a zeta function. Note that for $k=-1$  the  zeta function is zero
and contributions come only from $k=0$ and $k=1$.

The integrations in $H^{as}_k$ can be done and result in gamma and hypergeometric functions, which immediately allow for the continuation to $s=0$.

\begin{figure}[t]
%file is C:\\Users\\bordag\\WORK\\Physics\\Vacuum, effPot, \
%Casimir25\\Casimir25\\250318 figE.nb
\includegraphics[width=0.49\textwidth]{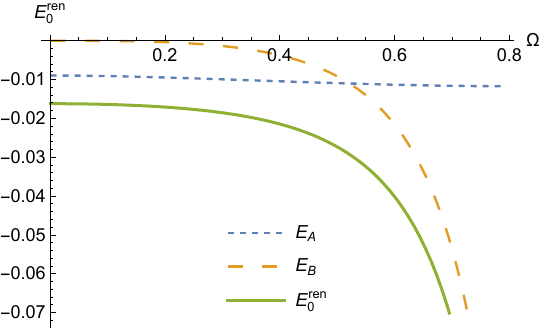}	\caption{
	The renormalized vacuum energy $E_0^{ren}$, \Ref{h.3}, with the added contributions of $E^{ren}_{l=0}$, \Ref{5.7}, and the contributions of $E_A$ and $E_B$,    \Ref{2.8}.}		\label{fig:figE}	
\end{figure}
	
The pole contribution, resulting from all $k$, is,
\eq{3.20a}{ E_A^{pole}=\left(-\frac{1}{512}-\frac{1}{32\pi}\right)\frac{1}{s}.
}
This result is in agreement with \cite{lese96-250-448} (our $s$ is $s/2$ in that paper).

The regular part of $E^{as}_A$, \Ref{3.12}, together with $E^{num}_A$, \Ref{3.11}, is shown in figure \ref{fig:figE}.
From the above expressions \Ref{3.17}, entering $E_A^{as}$, \Ref{3.16}, it is easy to get the expansion for $\Om\to0$ up to $\Om^2$,
\eq{3.21}{E_A^{as}=const+\frac{(\pi -12) (1+2 \pi )}{384 \pi ^2}\Om^2+O(\Om^4).
}

\section{Part B of the vacuum energy, $E_B$}
% see  250314 part B
In this section, we calculate part B of the vacuum energy defined in \Ref{2.8}. The branch point $\lambda_2$ on the positive real axis exists only for positive $l$.  We notice that a sign change $\Om\to-\Om$ makes it necessary to change $l\to-\l$ for part B to give a contribution. This way, this part of the energy does not depend on the sign of $\Om$ (for part A this is obvious) \textcolor{black}{and, consequently, it depends on $|\Om|$. In the rest of this section we take $\Om>0$ for conveneience and drop the module signs}. 
The segment of the contour that embraces the horizontal section of the cut, see figure \ref{fig:fig1}, contributes to the energy as follows,
\eq{4.1}{ E_B& =
	\frac12\sum_{l=1}^\infty \int_0^{\lambda_2} d\lambda
	\frac{ - \left(-e^{i\pi}g(\la)\right)^{1-2s}
		+\left(-e^{-i\pi}g(\la)\right)^{1-2s}}
	{2\pi i}     \nn   
	\\ & ~~~~~~~~~~~~~~~~~~~~~~~~~~~~~~~~~~~~\quad \quad \cdot \frac{\pa}{\pa\lambda}\ln\Phi(\lambda),
}
where $g(\la)$ is defined in \Ref{2.7a} and  $\lambda_2=\Omega l$ is the position of the branch point. In the massless case, $m=0$, we arrive at
\eq{4.3}{ E_B =-\frac{\sin(2 \pi s)}{2\pi}\sum_{l=1}^\infty \int_0^{\lambda_2} d\lambda
	\left(\Om l-\la\right)^{1-2 s} \ 
	\frac{\pa}{\pa\lambda}\ln\Phi(\lambda).
}
Following the algorithm of the previous section, we replace $\lambda=l z$ and note
\eq{4.5}{ E_B =-\frac{\sin(2 \pi s)}{2\pi}\sum_{l=1}^{\infty} l^{1-2s}
	\int_0^{\Omega } dz
	\left(\Om-z\right)^{1-2 s}  \ v(z),
}
where we use
\eq{4.6}{ v(z)= \frac{\pa}{\pa z}\ln\left(z^{-l}J_l(l z)\right),
}
and $J_l(l z)$ is the   Bessel function of first kind. 

The continuation in $s$ is obtained by subtracting and adding the function $v^{as}(z)$,
\eq{4.7}{v^{as}(z)=\sum_{k=-1}^2\frac{v^{as}_k(z)}{l^k},
}	
which is defined as the uniform expansion of $v(z)$ for large $l$ up to $l^{-2}$. It can be obtained from the Debye expansion of the Bessel function,
\eq{4.8}{ J_l(l z)=e^{l\eta(q)}\sqrt{\frac{q}{2\pi l}}
	\left(\sum_{k=0}^{\infty}\frac{U_k(q)}{l^k}\right),
}
where the coefficients $U_k(q)$ and the function $\eta(q)$ are the same as in \eqref{3.8}, with $p\to q$, $q=1/\sqrt{1-z^2}$.
% and we have included the prefactos in the coefficients.

With \Ref{4.7} we represent
\eq{4.9}{ E_B=E_B^{num}+E_B^{as}
}
with
\eq{4.10}{ E_B^{num} &=-\frac{\sin(2 \pi s)}{2\pi}\sum_{l=1}^\infty l^{1-2s} \int_0^\Omega dz
	\left(\Om-z\right)^{1-2s}   \
	\\\nn &~~~~~~~~~~~~~~~~~~~~~~~~~~~~~~~~~~~~~~\cdot (v(z)-v^{as}(z)),
	\\\nn E_B^{as}&= -\frac{\sin(2 \pi s)}{2\pi}\sum_{l=1}^\infty l^{1-2s} 
	\int_0^\Omega dz \left(\Om-z\right)^{1-2s} v^{as}(z).
}
In the numerical part, $E^{num}_B$, the integration over $z$ converges for $s=0$. The resulting integrals decrease sufficiently fast with increasing $l$ and the sum converges also for $s=0$.  Due to the factor $\sin(2 \pi s)$ the numerical part vanishes at $s=0$.

In the asymptotic part, all $z$-integrations converge for $s=0$ since these are over a finite range. Hence the only pole contribution which may compensate the sine in front, can come from the zeta function for $k=2$.
Accordingly, the only surviving term is
\eq{4.11}{E_B^{as}(s)
	&= -\frac{\sin(2 \pi s)}{2\pi} \zeta(1+2s) 
\\\nn &	~~~~~~\cdot\int_0^\Omega dz \left(\Om-z\right)^{1-2 s} v_2^{as}(z)+O(s),
}
with
\eq{4.12}{
	v_2^{as}(z)=\frac{z(4+10z^2+z^4)}{8(z^2-1)^4}.
}
The integration can be done for $s=0$ and we arrive finally at
\eq{4.13}{& E_B~=\lim_{s\to0}E_B^{as}(s)  \nonumber
\\&~~~~~~=-
	\frac{ \Omega}{256} 
\Bigg[\frac{1+9 \, \Omega^2}{\left(1-\Omega^2\right)^2}-
\frac{\mbox{ArcTan}\left(\Omega\right)}{\Omega}\Bigg],
}
where we have taken into account that $\sin(2 \pi s)\zeta(1+2s)|_{s\to0}=\pi$. The plot of this contribution to the vacuum energy is shown ~in Fig.\ref{fig:figE}. We mention that   $E_B$ has a finite continuation to $s=0$ and that therefore it does not have any pole contribution. However, its finite contribution is dominating, as can be seen from the figure. For small $\Om$ we get from \Ref{4.13} immediately
\eq{4.14}{E_B =-\frac{1}{24}\Om^3+O(\Om^5).
}

\subsection{The contribution of $l=0$}
%see file 250309_1 om=0 and m=0 and l=0 about zeros.nb
%
So far we have considered the rotation dependent contributions resulting from the orbital momenta $l\ne 0$ (except for \Ref{3.11}). In this section we consider for completeness the contribution of $l=0$ in \Ref{2.5}, even though it does not depend on the rotation. In this case, it is convenient to use the sum representation,
\eq{5.1}{ E_{l=0}=\frac12\sum_{n=1}^\infty {j_{0,n}}^{1-2s},
}
which follows from \Ref{2.5} and \Ref{2.4} for $m=0$ and $R=1$. To do the continuation to $s=0$, we use McMahon's expansion for large $n$, see eq. (10.21.(vi)) in \cite{NIST:DLMF}. It reads
\eq{5.2}{j_{l,n}=a-\frac{\mu-1}{8a}+\dots\,,
}
with $a=\pi\left(n+\frac{l}{2}-\frac14\right)$ and $\mu=4l^2$.  Defining $j_n^{as}$ from these two terms with $l=0$, we split the energy,
\eq{5.3}{E_{l=0}=E_{l=0}^{as}+E_{l=0}^{num}
}
with
\eq{5.4}{E_{l=0}^{num} = \frac12\sum_{n=0}^\infty \left( j_{0,n}-j^{as}\right),
	\qquad E_{l=0}^{as}= \frac12\sum_{n=0}^\infty (j_n^{as})^{1-2s}.
}
The sum in $E_{l=0}^{num} $ is convergent and returns a number, $E_{l=0}^{num}=-0.00256$. The continuation in $E_{l=0}^{as}$ gives a pole contribution,
\eq{5.5}{ E_{l=0,\ pole}^{as}=\frac{1}{32\pi s}+O(1),
}
which cancels out a corresponding contribution in $E_A^{pole}$, \Ref{3.20a}, and a regular contribution,
\eq{5.6}{E_{l=0,\ reg.}^{as}=\frac{1}{16\pi}\left(-1+\ga-\ln(\pi/8)\right)+\frac{\pi}{192}-\frac{1}{32}.
}
Together we get from $E_{l=0}^{num}$ and $E_{l=0}^{as}$ in \Ref{5.3} as finite contribution from $l=0$.
\eq{5.7}{E_{l=0}^{ren}=-0.007255.
}

\section{Conclusions}
In the preceding  section we calculated the vacuum energy for a scalar field rotating on a disk. We calculated separately the rotation-dependent parts $E_A$ and $E_B$, \Ref{2.8}, and the rotation-independent part $E_{l=0}$, \Ref{5.1}, which results from the mode with $l=0$. For the renormalized energy, we obtained an expression in terms of convergent sums/integrals, which allows an easy numerical evaluation. The results are shown in the figure \ref{fig:figE}.  It is noteworthy that this vacuum energy   is negative.
For small rotation frequency $\Om$, the expansion starts with $\Om^2$ with a negative coefficient (eq. \Ref{3.21}). The same holds for the next order in $\Om$, shown in \Ref{4.14}.  We mention that the vacuum energy must not depend on the sign of $\Om$ and that our calculations  are done for $\Om>0$. Hence the expansion of part B goes in fact in terms of $|\Om|$. \textcolor{black}{From the technical point of view, as we already mentioned, for $\Om<0$, orbital momenta with $l<0$ would contribute to part B. The result would be the same since $\Om$ and $l$ enter in the combination $\Om l$. Thus, the dependence of the vacuum energy on $\Om$ is not analytical and may be interpreted as a kind of anomaly.  Note, that this anomalous behavior is bound to the case $m=0$, which we considered here for the sake of simple calculations. For $m>0$, the potentially dangerous part B would be present for $\Om>\frac{m}{l}$ ($l=1,2,\dots$) and would not contribute to an expansion in $\Om=0$.
}

The renormalization is done by subtracting the pole part in the zeta function regularization. This pole part is expressed by eq. \Ref{h.2} in terms of the heat kernel coefficient $a_{3/2}$. In this approach we follow the scheme of section 4 in \cite{BKMM}. As is known, in the massless case there remains an ambiguity $\Delta E_0=\frac{a_{3/2}}{16\pi^{3/2}}\ln \mu^2$. Therefore our result, shown in figure \ref{fig:figE}, may be shifted by an arbitrary constant contribution.

The ultraviolet divergence of the vacuum energy ma\-nifested itself  in the pole contribution. This comes from part A, \Ref{3.20a} and the ($l=0$)-part, \Ref{5.5},
\eq{6.1}{E_0^{div}=E_A^{pole}+E^{as}_{l=0,\, pole}=-\frac{1}{512s}
}
From $E_B$ there was no pole contribution and $E^{as}_{l=0,\, pole}$ compensated a part of $E_A^{pole}$. It is remarkable that $E_0^{div}$ does not depend on the rotation. The corresponding heat kernel coefficient,
\eq{6.2}{a_{3/2}=-16\pi^{3/2} s E_0^{div}=\frac{\pi^{3/2}}{32},
}
is that known from the earlier calculations.

 The independence of the divergent part of the vacuum energy of the rotation reduces the interpretation of the renormalization to the previous, non-rotational case. The fact that the rotation-induced vacuum energy has no ultraviolet divergence is not surprising, since the ultraviolet divergences in QFT on a curved background are proportional to the curvature, while the metric \Ref{2.1} actually corresponds to a flat spacetime.

The calculation of the vacuum energy was performed by transforming the sum over $n$ in \Ref{2.5} into an integration in \Ref{2.7}, with the integration path shifted towards the imaginary axis, as illustrated in figure 1. This approach has the advantage of circumventing  oscillatory contributions, however, it may be technically more intricate to continue in the complex plane. Alternatively one may employ the original sum representation and the asymptotic expansion of the zeros or, more generally, of the eigenvalues $\om_n$. As an example, we demonstrated  this method for the calculation of the $l=0$ contribution, \Ref{5.1}, where it worked well. It is noteworthy that a similar method can be formulated for a smooth background field   in terms of the scattering phase shifts, as demonstrated in \cite{beau15-48-095401} for a simple example. However, for more complex fields,  the phase shifts may exhibit irregular behavior, as observed in \cite{bord18-51-455001}. Consequently, there is no universally applicable approach; rather, the method must be adapted to suit the specific problem at hand.

As can be seen from figure \ref{fig:figE}, the vacuum energy is negative. This property becomes particularly intriguing when contemplating the energy balance of the entire system, looking for a possible condensate. Further interest may result when incorporating a magnetic field, as previously  discussed in the introduction.  It is noteworthy that a recent study, \cite{bord25-40-2543018}, has examined the Casimir effect with a condensate.

A separate problem is the validity of the approximation introduced by the metric \Ref{2.6}. As previously mentioned, this approximation results from a Galilean transformation and does not conform to special relativity. This discrepancy is reflected in the restriction $\Om R<1$ and the need to introduce an artificial limit with some boundary condition. Consequently, one might expect the physical sense of the results to be restricted to $\Om R\ll 1$. However, the scalar field, being a relativistic object, generally inhabits the entire space, thereby imposing a significant restriction on the metric. One potential solution to this challenge could be the utilization of the relativistic metric following from \Ref{1.3} which was  proposed in \cite{fran22-8-265}. Nevertheless, this approach must be left for future investigation at this juncture.

\acknowledgments
We thank O.V. Teryaev and D.N. Voskre\-sensky for valuable and helpful discussions.

% 
% 
%\bibliographystyle{unsrt}
%\bibliography{C:/Users/Bordag/WORK/Literatur/bib/papers}

\begin{thebibliography}{10}
	
	\bibitem{fran22-8-265}
	Philip Franklin.
	\newblock {The Meaning of Rotation in the Special Theory of Relativity}.
	\newblock {\em Proceedings of the National Academy of Sciences of the United
		States of America}, 8(9):265--268, 1922.
	
	\bibitem{delo00-17-4241}
	V~A~De Lorenci, R~D M~De Paola, and N~F Svaiter.
	\newblock The rotating detector and vacuum fluctuations.
	\newblock {\em Classical and Quantum Gravity}, 17(20):4241, 2000.
	
	\bibitem{full09-42-155402}
	S.A. Fulling, L.~Kaplan, K.~Kirsten, Z.H. Liu, and K.A. Milton.
	\newblock {Vacuum Stress and Closed Paths in Rectangles, Pistons, and Pistols}.
	\newblock {\em J. Phys.}, A42:155402, 2009.
	
	\bibitem{cher1203.6588}
	M.~N. Chernodub.
	\newblock Permanently rotating devices: extracting rotation from quantum vacuum
	fluctuations?
	\newblock 2012.
	\newblock arXiv 1203.6588.
	
	\bibitem{cher13-87-025021}
	M.~N. Chernodub.
	\newblock Rotating casimir systems: Magnetic-field-enhanced perpetual motion,
	possible realization in doped nanotubes, and laws of thermodynamics.
	\newblock {\em Phys. Rev. D}, 87:025021, 2013.
	
	\bibitem{brag24-110-014511}
	Victor~V. Braguta, Maxim~N. Chernodub, Ilya~E. Kudrov, Artem~A. Roenko, and
	Dmitrii~A. Sychev.
	\newblock Negative barnett effect, negative moment of inertia of the gluon
	plasma, and thermal evaporation of the chromomagnetic condensate.
	\newblock {\em Phys. Rev. D}, 110:014511, 2024.
	
	\bibitem{brag24-852-138604}
	Victor~V. Braguta, Maxim~N. Chernodub, Artem~A. Roenko, and Dmitrii~A. Sychev.
	\newblock Negative moment of inertia and rotational instability of gluon
	plasma.
	\newblock {\em Physics Letters B}, 852:138604, 2024.
	
	\bibitem{liuz18-120-032001}
	Yizhuang Liu and Ismail Zahed.
	\newblock Pion condensation by rotation in a magnetic field.
	\newblock {\em Phys. Rev. Lett.}, 120:032001, 2018.
	
	\bibitem{guot22-106-094010}
	Tao Guo, Jianing Li, Chengfu Mu, and Lianyi He.
	\newblock {Formation of a supergiant quantum vortex in a relativistic
		Bose-Einstein condensate driven by rotation and a parallel magnetic field}.
	\newblock {\em Phys. Rev. D}, 106:094010, 2022.
	
	\bibitem{nour10-82-044047}
	M.~Nouri-Zonoz and B.~Nazari.
	\newblock Vacuum energy and the spacetime index of refraction: A new synthesis.
	\newblock {\em Phys. Rev. D}, 82:044047, 2010.
	
	\bibitem{beze14-89-044015}
	V.~B. Bezerra, H.~F. Mota, and C.~R. Muniz.
	\newblock Casimir effect due to a slowly rotating source in the weak-field
	approximation.
	\newblock {\em Phys. Rev. D}, 89:044015, 2014.
	
	\bibitem{BKMM}
	M.~Bordag, G.~L. Klimchitskaya, U.~Mohideen, and V.~M. Mostepanenko.
	\newblock {\em Advances in the Casimir Effect}.
	\newblock Oxford University Press, Oxford, 2009.
	
	\bibitem{bord01-353-1}
	M.~Bordag, U.~Mohideen, and V.~M. Mostepanenko.
	\newblock {New Developments in the Casimir Effect}.
	\newblock {\em Phys. Rept.}, 353:1--205, 2001.
	
	\bibitem{AbramowitzStegun2010}
	F.W.J. Olver, D.W. Lozier, R.F. Boisvert, and Ch.W. Clark.
	\newblock {\em NIST Handbook of mathematical functions: with formulas, graphs,
		and mathematical tables}.
	\newblock Cambridge university press, 2010.
	
	\bibitem{NIST:DLMF}
	{\it NIST Digital Library of Mathematical Functions}.
	\newblock {https://dlmf.nist.gov/}, Release 1.2.3 of 2024-12-15.
	
	\bibitem{lese96-250-448}
	S~Leseduarte and A~Romeo.
	\newblock {Complete zeta-function approach to the electromagnetic Casimir
		effect for spheres and circles}.
	\newblock {\em Annals of Physics}, 250(2):448--484, 1996.
	
	\bibitem{beau15-48-095401}
	Matthew Beauregard, Michael Bordag, and Klaus Kirsten.
	\newblock {Casimir energies in spherically symmetric background potentials
		revisited}.
	\newblock {\em J.~Phys.~A: Math.~Gen.}, 48:095401, 2015.
	
	\bibitem{bord18-51-455001}
	M~Bordag and K~Kirsten.
	\newblock On the entropy of a spherical plasma shell.
	\newblock {\em J.~Phys.~A: Math.~Gen.}, 51:455001, 2018.
	
	\bibitem{bord25-40-2543018}
	M.~Bordag and I.~G. Pirozhenko.
	\newblock {Casimir effect with an unstable mode}.
	\newblock {\em Int. J. Mod. Phys. A}, 40(10/11):2543018, 2025.
	
\end{thebibliography}
% 	

\end{document}